%% file: holdspec.tex
\documentclass[conference]{IEEEtran}
\IEEEoverridecommandlockouts

\usepackage{cite}
\usepackage{amsmath,amssymb}
\usepackage{booktabs}
\usepackage{graphicx}
\usepackage{listings}
\usepackage{xcolor}
\usepackage{url}
\usepackage[hidelinks]{hyperref}

\input{macros}

\newcommand{\code}[1]{\texttt{\small #1}}

\begin{document}

\title{HoldSpec: A Machine-Checked Model and Conformance Suite\\
for the Payment Authorization Hold Lifecycle}

\author{\IEEEauthorblockN{Abhishek Sharma}
\IEEEauthorblockA{Senior Member, IEEE\quad\textbar\quad ORCID 0009-0007-1103-2103\\
abhicse24@gmail.com\quad\textbar\quad github.com/abhisheksharma2411\quad\textbar\quad abhisharma.co.in}}

\maketitle

\begin{abstract}
When a merchant authorizes a card and captures the funds later, the money sits
in a state that neither party fully controls: held on the cardholder's account,
not yet taken, and due to lapse on a deadline set by the card network. What a
merchant may do to that hold, capture part of it, capture again, void it,
raise it, let it expire, is business logic that every payment service provider
implements slightly differently and none of them specifies formally. We give the
lifecycle a state machine in TLA+, parameterized by a provider capability
profile, and check eight safety invariants and three liveness properties over
\numprofiles{} profiles, five of them built from Stripe's and Adyen's published
documentation.
From the same machine we generate a conformance suite that runs black-box
against a provider API. Against \killablemutants{} seeded defects that are
distinguishable from a correct implementation, the generated suite detects
\modelrate{}\%, where random sequences with the same call budget reach
\randomrate{}\% and a hand-written integration suite reaches \handrate{}\%;
random sequences given ten times the budget still reach only
\randomtenxrate{}\%. Running the same suite between two providers turns the
comparison into a search, and it returns short executable witnesses for
divergences the two documentation sets imply but neither states, including
one that breaks the common assumption that a single adapter interface can sit
over both, because closing a hold and releasing its remainder is a capture on
one provider and a cancel on the other. We report what we could not do: no
provider sandbox credentials were available, so every result here concerns
documented behavior, and the suite that would test the providers themselves is
part of what we release.
\end{abstract}


\section{Introduction}

A hotel takes a card at check-in and charges it at check-out. A retailer
authorizes an order and captures each parcel as it ships. A car rental firm holds
an estimate and settles the real amount days later. All three use the same
mechanism: an authorization that reserves funds without moving them, followed by
one or more captures that move some or all of what was reserved, and a deadline
after which whatever was not captured goes back.

The mechanism is old and the rules around it are not written down anywhere a
machine can read. Each payment service provider (PSP) documents its own version
in prose: how long an authorization stays good, whether a second capture is
possible, whether more than the authorized amount may be taken, what happens to
the remainder after a partial capture. The rules differ between providers, they
differ between accounts at the same provider, and the differences are the kind
that only surface in production, on somebody's money.

The defects are correspondingly unglamorous and expensive. A capture that
arrives after a void. A retry loop that captures twice against one
authorization. An over-capture that a provider silently accepts and a card
network later charges back. A hold that closes without the funds being released,
so a customer's balance stays short for a week. Each is a violation of a rule
that a state machine could state in one line, and that no PSP states in a form
anything can check.

This paper writes that state machine down and then puts it to work.

\subsection{Contributions}

\begin{itemize}
\item \textbf{A formal model of the hold lifecycle} (\S\ref{sec:model}). A TLA+
  specification of authorization, partial and multiple capture, void, expiry,
  incremental authorization, and the release of held funds, parameterized by a
  provider capability profile. Eight safety invariants and three liveness
  properties are checked exhaustively over \numprofiles{} profiles;
  \totaldistinctstates{} distinct states in total, in \checkseconds{} seconds.

\item \textbf{Evidence that the checking is not vacuous} (\S\ref{sec:vacuity}).
  \numspecmutations{} targeted changes to the specification, one per property.
  Every property is falsified by the change aimed at it, and no change survives
  every property.

\item \textbf{A conformance suite generated from the model}
  (\S\ref{sec:conformance}), executable black-box over HTTP. On
  \killablemutants{} seeded defects it detects \modelrate{}\%, against
  \randomrate{}\% for equal-budget random testing and \handrate{}\% for a
  hand-written suite. Defects that no call sequence can distinguish from correct
  behavior are identified by exhaustive search and reported separately rather
  than counted as misses.

\item \textbf{A differential procedure for comparing providers}
  (\S\ref{sec:divergence}) that enumerates, exhaustively within the model's
  bounds, every way two profiles disagree, with the shortest call sequence that
  shows each one. Applied to Stripe and Adyen it returns divergences in
  authorization validity, over-capture, and the mechanism for releasing a
  remainder.

\item \textbf{Two negative results about observability}
  (\S\ref{sec:observability}) that we did not expect and that constrain any
  conformance work in this area: neither the number of captures nor the reason a
  hold closed is a provider-neutral observable, and the second is not
  synchronously observable at all.
\end{itemize}

Everything is released under MIT: the specification, the model checker
configurations, the suite, the mock providers, the sandbox adapters, and the
scripts that regenerate every number and figure here.

\section{The hold lifecycle}
\label{sec:background}

\subsection{What a hold is}

An authorization asks the issuer to reserve an amount on a card. The issuer
answers yes or no, and on yes the amount stops being available to the cardholder
without yet belonging to the merchant. Capture is the separate step that moves
it. Between the two, three things can happen: the merchant captures (possibly
less than was authorized, possibly more than once, possibly more than was
authorized), the merchant voids, or the clock runs out.

The rules are not the merchant's to choose. Card networks set the validity
window, and a capture after it may be rejected or may incur a fee. The
provider's API surface then decides which of the network's possibilities the
merchant can actually reach.

\subsection{Why this is not idempotency}

Duplicate-effect prevention is the neighboring problem and a different one.
Idempotency is about one operation submitted more than once; the hold lifecycle
is about one authorization over its life. An implementation can be flawlessly
idempotent and still accept a capture after a void, because the capture is not a
retry of anything, it is a first attempt at an operation that should no longer
be legal. It can be correct about holds and still double-charge on a network
retry. The two are independent, and we treat them separately: the idempotency
contract is the subject of a companion manuscript~\cite{sharma2026idempotency},
and nothing here depends on it.

\subsection{What the providers document}
\label{sec:providers}

Table~\ref{tab:profiles} gives the profiles used throughout, each field taken
from published documentation with the URL and the quotation recorded in the
artifact. Four points are worth drawing out because the experiments turn on
them.

Stripe permits one capture on most payments; a partial capture releases the
remainder automatically~\cite{stripe2026hold}. With multicapture, up to 50
non-final captures are allowed followed by one final capture, and
\code{final\_capture=false} keeps the remaining amount
authorized~\cite{stripe2026multicapture}. With overcapture, a capture may exceed
the authorized amount by a network- and category-dependent margin, generally
15\% and up to 30\% for some merchant categories~\cite{stripe2026overcapture}.

Adyen's default is a single partial capture after which ``[a]ny unclaimed amount
that is left over after partially capturing a payment is automatically
cancelled''; enabling multiple partial captures changes that, so ``[t]he
unclaimed amount after an initial partial capture is not automatically
cancelled''~\cite{adyen2026capture}. A capture amount ``must be the same as or,
in case of a partial capture, less than the authorized amount'', no
over-capture. Adyen's validity table gives 10 days for Visa card-not-present
cardholder-initiated transactions~\cite{adyen2026adjust}, where Stripe documents
7 days for the same brand~\cite{stripe2026hold}.

Two structural differences follow from this and matter more than any single
number. Stripe's API has a flag that says ``this capture is the last one'';
Adyen's has no such flag, so on an account with multiple partial captures the
merchant ends a hold either by capturing the whole authorized amount or by
cancelling what is left. And where Stripe releases a remainder with a zero-amount
capture, Adyen releases it with a cancel.

\begin{table*}[t]
\centering
\caption{Provider profiles, from published documentation (accessed 1 September
2026). The synthetic profile exercises incremental authorization, which neither
provider's documented default includes.}
\label{tab:profiles}
\small
\input{tab_profiles}
\end{table*}

\section{The model}
\label{sec:model}

\subsection{State and actions}

A hold is described by eleven variables: its \code{state}
(\code{NONE}/\code{HELD}/\code{CLOSED}), how it closed, the authorized amount,
the captured total, the captured total recorded at the moment of closing, the
number of accepted captures, the time of the last capture, a count of
funds-release events, the expiry instant, the release deadline, and a clock.
Amounts are integer minor units; time is in ticks.

Eight actions move it. \code{Authorize} creates the hold.
\code{CaptureNonFinal} takes part of it and leaves the rest reserved.
\code{CaptureFinal} takes an amount and closes the hold, releasing whatever is
left; an amount of zero is the documented way to release a remainder without
taking more. \code{Void} closes an open hold. \code{Expire} closes it at its
deadline. \code{ReleaseHold} returns the reserved funds to the cardholder.
\code{IncreaseAuth} raises the authorized amount. \code{Tick} advances the clock.

Keeping \code{ReleaseHold} separate from the actions that close a hold is what
makes the interesting properties interesting. If closing a hold released its
funds in the same step, ``released exactly once'' and ``released promptly'' would
be true by construction and worth nothing. Split apart, they are claims about
the implementation that a suite can actually test, and, as
\S\ref{sec:conformance} shows, the mutants that break them are exactly the ones a
hand-written suite tends to miss.

\subsection{Profiles as parameters}

Everything provider-specific is a constant: the authorized amount and its
ceiling, the over-capture allowance, the non-final capture budget, whether
incremental authorization is offered, whether a final-capture flag exists,
whether a void is accepted after a capture, the validity window, and the release
deadline. One specification therefore covers every profile in
Table~\ref{tab:profiles}, and a provider difference is a difference in
constants rather than in code. This is what makes \S\ref{sec:divergence}
possible.

\subsection{Time}

Two deadlines behave differently, and the difference is deliberate. An
authorization expires at its deadline: the clock cannot advance past
\code{expiresAt} while a hold is open, so \code{Expire} fires there. Releasing
the funds is allowed to lag, but only by the profile's release delay, and once
that is exceeded the clock is blocked again. Both use the upper-bound-timer
treatment of real time in TLA+~\cite{abadi1994oldfashioned}: an action with a
deadline blocks time rather than being quietly missed, so a missed deadline
becomes a timelock that the model checker reports.

This was not the first design. Initially the clock advanced freely and
\code{INV\_BoundedRelease} was left as an invariant to check. TLC's first run
produced a four-step counterexample in which a hold closed, time simply ran on,
and the release never became due, correctly reporting that the specification
permitted what it was meant to forbid.

\subsection{Properties}

Eight safety invariants:
captures never exceed the profile's ceiling;
nothing is captured after a hold closes;
funds are released at most once;
funds are not released before the hold closes;
a closed hold releases within its deadline;
no capture is accepted at or after the expiry instant;
the capture-count budget is respected;
and a released hold leaves nothing held.
Three liveness properties, under weak fairness on \code{Authorize},
\code{Tick}, \code{Expire} and \code{ReleaseHold}: a closed hold eventually
releases; an open hold does not stay open forever; every behavior reaches a
settled end state.

\subsection{Model checking}

Table~\ref{tab:modelcheck} gives TLC's results. All eleven properties hold for
all \numprofiles{} profiles. The state spaces are small, the largest is
\largeststatespace{} distinct states, which is the point of a bounded model:
exhaustive checking finishes in \checkseconds{} seconds in total, so the
specification can be re-checked on every change to it, as
Howard~et~al.~\cite{howard2025ccf} argue specifications should be.

\begin{table}[t]
\centering
\caption{TLC results per profile. Eight invariants and three liveness properties
were enabled for every run.}
\label{tab:modelcheck}
\scriptsize
\input{tab_model_check}
\end{table}

\begin{figure}[t]
\centering
\includegraphics[width=\columnwidth]{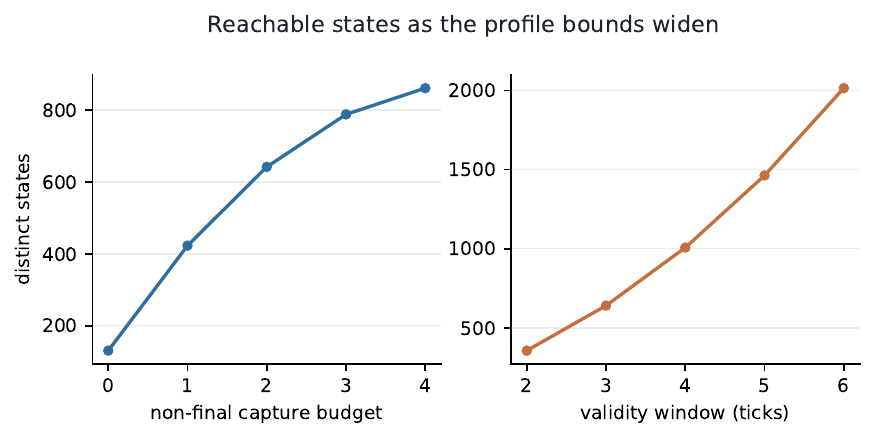}
\caption{The reachable space grows roughly linearly in the capture budget and
faster in the validity window, since a longer window multiplies the states in
which each capture can happen. Both stay small enough for exhaustive checking
across the range a provider profile needs.}
\label{fig:statespace}
\end{figure}

\section{Is the checking vacuous?}
\label{sec:vacuity}

A specification that satisfies its invariants because nothing can happen in it
is worthless, and the failure mode is quiet. To test for it we broke the
specification \numspecmutations{} times, once per property, and re-ran TLC.

Each mutation removes or weakens one guard: a capture that ignores the
authorization deadline, a capture that ignores the ceiling, a release that can
happen twice, a release that can happen while the hold is open, a clock that
runs past an overdue release, a capture budget that is not enforced, a close that
forgets what had been captured, and the removal of each of the three fairness
conditions.

Attribution needs care. TLC stops at the first property that fails, so with
everything enabled a mutation can be reported against a property it was not
aimed at, both ceiling mutations trip \code{TypeOK} first, and two of the
three fairness mutations trip \code{EventualRelease} before their own property.
Each mutation is therefore run twice: once with all properties, and once with
only the property it targets. Table~\ref{tab:specmutation} reports the second.
All \specmutationscaught{} are caught by the property aimed at them, and
\specmutationssurvived{} survive every property.

Two limits on what this shows. It establishes that each property has force, not
that the set is complete: a rule we never wrote down cannot fail here. And the
profile capability guards are not covered, because they are not safety
properties, weakening one makes the model describe a different provider rather
than an incorrect one. Measuring those is what \S\ref{sec:divergence} does.

\begin{table*}[t]
\centering
\caption{Specification mutations. ``Caught'' is with only the targeted property
enabled. Trace is the counterexample length TLC reported.}
\label{tab:specmutation}
\small
\input{tab_spec_mutation}
\end{table*}

\section{From model to conformance suite}
\label{sec:conformance}

\subsection{Keeping the two models in step}

The specification is checked by TLC, but TLC cannot generate test sequences or
grade a running service, so the state machine is also implemented in Python. Two
models of the same thing is an obvious way to be wrong slowly.

We check it directly: for each profile, both state spaces are enumerated and
compared element by element, using TLC's state dump. They are identical for all
\equivalenceprofiles{} profiles. This runs as an experiment rather than a
one-time check, so any future edit that moves one and not the other fails
immediately.

\subsection{What a test may do and see}

A test can call four operations, authorize, capture (with an amount and a
final flag), void, increase authorization, and can let time pass. It cannot
invoke expiry or the release of funds; those are the provider's to do, and the
test only waits and looks.

What it sees is the hold's status, the authorized amount, the captured total,
and whether the held funds have been released. The oracle asks the model, before
each call, whether a conforming provider must accept it and what must be
observable afterwards, then compares. Release is checked against a permitted set
rather than a value, since the profile allows it to lag: while the deadline has
not passed either answer is conforming, and after it only one is.

\subsection{Two things that turned out not to be observable}
\label{sec:observability}

Two fields were in the observation at first and had to come out. Both removals
are findings rather than concessions.

\emph{The number of captures is not provider-neutral.} Closing a hold and
releasing its remainder costs one capture call on Stripe, where it is a
zero-amount capture with \code{final\_capture=true}, and zero on Adyen, where it
is a cancel. A suite that counts captures is counting different things on the
two rails. This surfaced as a conformance failure in the HTTP run, where the
Adyen adapter's translation of one abstract operation into two requests showed up
as an extra capture.

\emph{The reason a hold closed is not observable at all, synchronously.} A
Stripe PaymentIntent that was voided and one that expired uncaptured both read
\code{canceled}~\cite{stripe2026hold}; Adyen reports a cancelled and an expired
authorisation alike. Distinguishing them requires the event stream. We removed
the field rather than grade providers on something their APIs do not answer.

Both removals cost nothing measurable, detection stayed at
\modeldetected{}/\killablemutants{}, but that is evidence about these
particular defects, not a proof that nothing needs those fields. A defect that
voids a hold where it should have let it expire is invisible to this suite, and
consuming the event stream is the way to reach it.

\subsection{Generating the suite}

The suite covers every state a black-box test can drive the provider into, every
accepted transition between those states, and, at every state, every call the
model says must be refused. Construction is a breadth-first tree over the
testable state space: one test walks each root-to-leaf path, and each transition
the tree does not use gets one short test. Because the oracle checks after every
call, a test that is a prefix of another adds nothing and is dropped. Refusal
checks are issued at each state along the way, which is cheap: a refused call
leaves the state alone, so they can all be made at one point in a sequence. The
criterion is standard for testing from state machines~\cite{chow1978testing,
lee1996principles}; what is new is the machine it is applied to.

\subsection{Seeded defects}

Twelve mutants, each breaking one rule, in the style of mutation
testing~\cite{demillo1978hints, jia2011analysis}: capture after close, capture
after expiry, an unbounded ceiling, a ceiling off by one, an extra non-final
capture, a double release, no release, a late release, a void accepted after a
capture where the profile forbids it, a partial capture that leaves the hold
open, an expiry check off by one tick, and a release before the hold closes.
Each overrides a single decision in the reference provider, so a mutant differs
from a correct implementation by one choice rather than a rewrite.

Not every mutant is detectable. On a single-capture profile, ``void accepted
after a capture'' cannot be reached, because any capture closes the hold first.
Rather than assume which are which, we search: a breadth-first walk over the
product of mutant and reference under identical call sequences, returning the
shortest distinguishing sequence or, exhaustively within the profile's bounds,
nothing. Mutants in the second category are reported as equivalent and excluded
from the rates. Of 72 mutant--profile pairs, \equivmutants{} are equivalent and
\killablemutants{} are killable.

\subsection{Baselines}

Two, both using the same oracle, so what is compared is the choice of call
sequences and nothing else. \emph{Random} draws calls uniformly, with the same
API-call budget as the generated suite, refusal checks included, since those
cost a call too. A second random arm gets ten times that budget, to separate
choosing the wrong calls from not making enough of them. \emph{Hand-written} is
the suite an integration engineer writes from a provider quickstart: authorize
then capture, authorize then void, authorize and let it expire, a partial
capture, a capture over the limit, a second capture, a capture after a void, a
capture with no authorization, and the profile-specific happy path. Eight to ten
tests, no exhaustive refusal checking, which is the point of it.

\section{Results}
\label{sec:results}

\subsection{Detection}

\begin{figure}[t]
\centering
\includegraphics[width=\columnwidth]{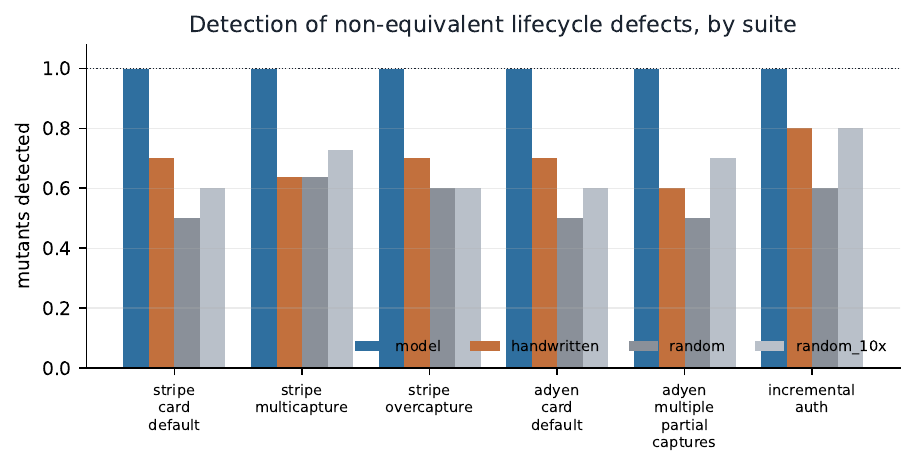}
\caption{Detection by suite and profile. The generated suite is at 1.0
everywhere; the baselines vary with which defects happen to lie on a common
path.}
\label{fig:detection}
\end{figure}

Table~\ref{tab:conformance} and Fig.~\ref{fig:detection} give the outcome. The
generated suite detects every killable defect on every profile,
\modeldetected{}/\killablemutants{}.

That figure should be read with its construction in mind. A defect counts as
killable when some sequence within the model's bounds distinguishes it, and the
generated suite is a breadth-first enumeration of sequences within those same
bounds. Complete detection is close to structurally guaranteed and is not the
interesting number; it confirms that the generator covers the space its own
killability criterion is defined over. What the experiment does measure is how
far the baselines fall short of that coverage under an identical call budget,
and both are scored against the same killability criterion, so the comparison
between them is fair.

Equal-budget random reaches \randomrate{}\%, and with ten times the budget
\randomtenxrate{}\%, so the gap is in which sequences are chosen rather than how
many. The hand-written suite reaches \handrate{}\% on roughly 25 calls, a
respectable return for its size: it is a cheap baseline rather than a weak one.
With \killablemutants{} defects in the denominator, a single detection is worth
about \pointspermutant{} percentage points, so the two-point difference between the
hand-written suite and random at ten times the budget should not be read as an
ordering.

What the baselines miss is consistent and, in hindsight, predictable. They find
the defects that break a common path, an unbounded ceiling shows up the first
time anything captures too much. They miss the ones that need a specific state
reached in a specific order: a capture accepted after a void needs a void first
and a capture second, both on a hold that is otherwise fine; a late release needs
the clock pushed past a deadline that only exists after the hold closes. The
generated suite reaches those because it is built from the state space rather
than from a list of scenarios someone thought of.

\begin{table*}[t]
\centering
\caption{Detection of seeded defects. Equivalent mutants, those no call
sequence can distinguish from a correct implementation, established by
exhaustive search, are excluded from the rates. Random arms use the model
suite's call budget and ten times it.}
\label{tab:conformance}
\small
\input{tab_conformance}
\end{table*}

\subsection{Black-box over HTTP}

An in-process result proves less than it appears to, because the suite and the
system under test share a language and a process. We therefore stood up mock
providers that speak Stripe's and Adyen's own request shapes over HTTP ---
\code{POST /v1/payment\_intents/\{id\}/capture} with
\code{amount\_to\_capture} and \code{final\_capture} on one,
\code{POST /payments/\{ref\}/captures} with an amount object on the other, and
ran the identical suite through adapters.

Table~\ref{tab:http} reports it. All \httpconforming{} of \httpdeployments{}
deployments conform, and all \httpinjecteddetected{} of \httpinjected{} injected
defects are detected through the network interface.

The value of this run was not the confirmation. It was the failures along the
way: the Adyen adapter did not conform at first, and each failure was a real
defect in the model rather than in the adapter. That is how the missing
final-capture capability was found (\S\ref{sec:providers}), and how both
non-observable fields were found (\S\ref{sec:observability}). A specification
written only against documentation and checked only against itself would have
kept all three mistakes.

\begin{table}[t]
\centering
\caption{The same suite over HTTP, against services speaking each provider's
request shape.}
\label{tab:http}
\scriptsize
\input{tab_http}
\end{table}

\subsection{Where two providers disagree}
\label{sec:divergence}

The same product search that classifies equivalent mutants compares two
providers when the two sides are given different profiles. Run to exhaustion it
returns every reachable disagreement; divergences reached by different prefixes
but making the same attributes differ in the same way are one finding, and the
shortest prefix is kept as the witness.

Two adjustments were needed to make the answers mean anything. Both sides get
the larger of the two clock bounds, because otherwise the profile with the
shorter horizon simply stops advancing and the saturation is reported as a
behavioral difference, before this was fixed, three of four reported
divergences in the Stripe/Adyen comparison were that artifact. And a class is
identified independently of the amount, because a provider that refuses a closing
capture refuses it at every amount, and reporting one class per amount reports
one disagreement several times. The amounts are kept alongside the class, since
for a bound like over-capture the amount at which behavior changes is the
finding.

\begin{figure}[t]
\centering
\includegraphics[width=0.86\columnwidth]{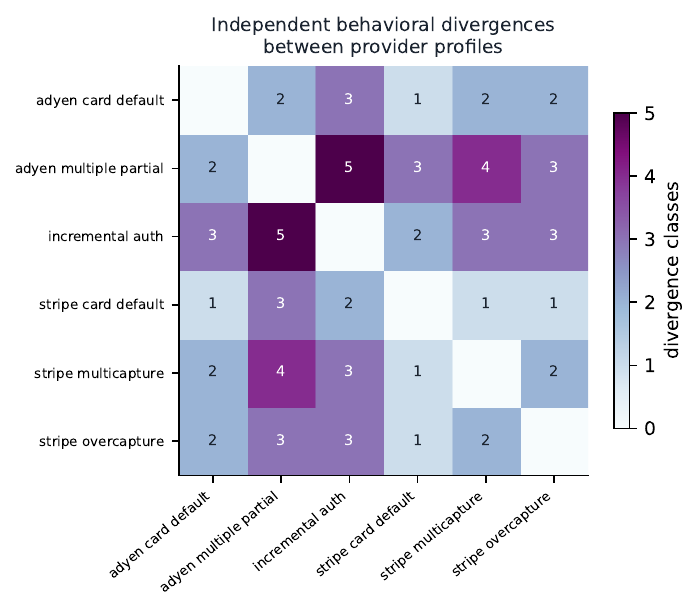}
\caption{Independent divergence classes between every pair of profiles. No pair
agrees everywhere, including the two Stripe profiles that differ only in an
account setting.}
\label{fig:divergence}
\end{figure}

Across all \numpairs{} profile pairs, every pair diverges, with up to
\maxdivergences{} independent classes (Fig.~\ref{fig:divergence}).
Table~\ref{tab:divergences} gives the three pairs that correspond to real
provider comparisons.

\emph{Defaults.} One divergence: the authorization expires a tick earlier on the
Stripe profile. This is the documented 7-day versus 10-day validity for Visa
card-not-present cardholder-initiated transactions, and the witness is a capture
attempt in the window where one provider still holds the funds and the other has
released them.

\emph{Over-capture.} Two: a capture of five units against an authorization of
four is accepted on the Stripe overcapture profile and refused on Adyen, plus
the validity difference. The witness is two calls long.

\emph{Multiple captures.} Four, and these are the ones that matter for anyone
building over both. A closing capture below the full authorized amount is
accepted on Stripe and refused on Adyen, because Adyen has no final-capture flag.
A non-final capture of the whole authorized amount is accepted on Stripe, which
keeps the PaymentIntent open with nothing left to capture, and refused on Adyen,
where taking everything ends the hold. A void after a partial capture is refused
on Stripe and accepted on Adyen, the same operation, opposite answers. And the
validity difference again.

\begin{table*}[t]
\centering
\caption{Divergences between provider profiles, with the shortest call sequence
that shows each one. In the witnesses \code{auth} is authorize, \code{wait} is
one tick of elapsed time, \code{cap($n$,fin)} is a capture of $n$ that closes
the hold and \code{cap($n$,part)} one that does not.}
\label{tab:divergences}
\scriptsize
\input{tab_divergences}
\end{table*}

\subsection{A reproducible defect corpus}

Every conformance run and every violation is recorded with the exact call
sequence that produced it, in Postgres when one is configured and in SQLite
otherwise. The runs reported here left \corpusrows{} reproducible violations. A
violation is thus a script rather than a log line: re-running it reproduces the
failure against the same provider profile.

\section{Discussion}

\subsection{One adapter over two rails is harder than it looks}

The usual way to support several PSPs is an internal interface, authorize,
capture, void, with an adapter behind each provider. The divergences above say
what that interface costs.

Closing a hold short of the full amount is a capture on Stripe and, on Adyen, not
expressible as a capture at all. Releasing a remainder is a zero-amount capture
on one and a cancel on the other. A void after a partial capture is legal on one
and refused on the other. An adapter can hide the first two by composing calls,
as ours does, but composition has a visible cost: what was one operation becomes
two, and any observable that counts operations stops meaning the same thing.
The third cannot be hidden, the operation is simply unavailable on one rail,
and the plain adapter refuses it locally rather than sending a request that will
be misinterpreted.

The practical reading: a multi-PSP interface should be typed by capability, and
the capability vector should be checked, not assumed. That is what a profile is,
and checking it is what the suite does.

\subsection{What the specification is for}

The model is small: eleven variables, eight actions, a few thousand states. Its
value is not that it is difficult but that it is checkable, and that everything
downstream is derived from it rather than restated alongside it. The suite is
generated from it, the oracle is it, the provider comparison is it under two
parameterizations. When the model was wrong, three times, all found by running
the suite against something that did not share its assumptions, one edit fixed
the specification, the suite, and the comparison together.

\section{Limitations}
\label{sec:limitations}

\emph{No live provider was tested.} No sandbox credentials were available. Every
result concerns behavior the providers document, not behavior they exhibit.
Whether Stripe and Adyen honor their own documentation is the question this work
was built to ask and has not answered. The adapters that would ask it are
released, unrun and marked as such.

\emph{Two obstacles stand between the suite and a live run}, beyond credentials.
A real authorization cannot be fast-forwarded, so expiry tests either wait days
or are excluded. And Adyen reports modification outcomes by webhook rather than
in the response, so a live run needs a notification endpoint; our adapter raises
rather than pretending to poll.

\emph{The model is bounded and small.} Four minor units, a handful of ticks, one
hold at a time. Defects that depend on magnitude, an overflow, a currency with
a different minor-unit scale, are out of reach, as are races between
concurrent operations on one authorization, which need the operation identity
that the idempotency work supplies.

\emph{``Equivalent'' means equivalent within those bounds.} A mutant the search
calls indistinguishable might be distinguishable at an amount or a horizon the
search does not reach.

\emph{Profiles are one reading of the documentation.} They were built by one
author from published pages, quoted and cited so they can be checked, but a
misreading would propagate silently into every result that depends on it.

\emph{Detection rates are about these twelve defects.} They are drawn from
defects we have seen in practice and they cover each invariant, but a defect
class nobody thought of is not in the denominator.

\section{Related work}

\emph{Formal methods for payments} have concentrated on the cryptographic
protocol. Basin~et~al.~\cite{basin2021emv} built a symbolic Tamarin model of EMV
and found two attacks on contactless transactions. That work is about
card--terminal authentication at the point of sale; the business logic that
governs the resulting authorization afterwards, partial capture, void, expiry
--- is outside its scope and is the subject here.

\emph{TLA+ applied to payment systems} is closest in method. Grundmann and
Hartenstein~\cite{grundmann2023lightning} verify the Lightning Network with a
refinement chain to keep the state space tractable, and their practice of
reporting state counts and depth rather than asserting a model was checked is one
we follow. The domain is different: payment channels, not merchant-side holds.

\emph{Binding specifications to implementations.} Howard~et~al.~\cite{howard2025ccf}
validate traces of the Confidential Consortium Framework against a TLA+
specification in CI and find six bugs. This is the model for
\S\ref{sec:conformance} and the reason we treat model-versus-implementation
agreement as an experiment rather than an assumption. Their system's source is
available, so they can validate traces; ours is a third-party API, so the binding
has to be black-box conformance. Newcombe~et~al.~\cite{newcombe2014amazon} make
the broader industrial case for specifying before building.

\emph{Differential testing of APIs.} Godefroid~et~al.~\cite{godefroid2020differential}
compare versions of a REST service to find breaking changes, driven by the API's
interface description. We compare two providers of the same business capability,
driven by a semantic state machine, so a divergence comes back as a lifecycle
rule rather than a schema change.

\emph{Testing from state machines} is long-settled~\cite{chow1978testing,
lee1996principles}, as is mutation testing~\cite{demillo1978hints,
jia2011analysis}. We use both as tools. The contribution is the machine they are
applied to and what applying them reveals about the providers.

\emph{Provider documentation}~\cite{stripe2026hold, stripe2026multicapture,
stripe2026overcapture, adyen2026capture, adyen2026adjust} is careful and, within
each provider, complete. None of it is machine-checkable, and none of it
describes another provider, which is precisely the gap
\S\ref{sec:divergence} fills.

\section{Conclusion}

The authorization hold is a piece of financial state that sits between two
parties for days, governed by rules that every payment provider implements and
none specifies formally. We gave those rules a machine-checked state machine,
parameterized it by provider capability, and used it to generate a conformance
suite that detects every seeded defect a call sequence can reach, where random
testing at ten times the budget reaches two-thirds and a hand-written suite about
the same.

The result we did not expect is how much the two providers differ under the same
name. Every profile pair diverges. Two of the divergences make the common
one-interface-over-many-rails design harder than it is usually treated as being,
and one of them, the absence of a final-capture flag, is not a difference in
values but in what operations exist.

The obvious next step is the one we could not take: point the suite at a live
sandbox and find out whether the providers do what they say. Everything needed to
do that is released.

\section*{Artifact Availability}

Code, specification, conformance suite, mock providers, sandbox adapters,
experiment scripts, and the scripts that regenerate every table and figure are
at \url{https://github.com/abhisheksharma2411/holdspec}, MIT licensed, tagged
\code{v1.0.4}, which is the revision this paper was built from. \code{make reproduce} runs the full pipeline from a clean tree:
model checking, all six experiments, figures, and this PDF. Total runtime is a
few minutes on a laptop; Docker is optional. The archived release is deposited on
Zenodo under DOI \texttt{\zenodocode}. There is no separate dataset: provider
profiles are version-controlled with a source URL and a verbatim quotation per
field, and all workloads are generated from the model at run time.

\bibliographystyle{IEEEtran}
\bibliography{refs}

\end{document}

%% file: macros.tex
\newcommand{\numprofiles}{6}
\newcommand{\totaldistinctstates}{4,096}
\newcommand{\largeststatespace}{2,248}
\newcommand{\checkseconds}{4.4}
\newcommand{\numspecmutations}{10}
\newcommand{\specmutationscaught}{10}
\newcommand{\specmutationssurvived}{0}
\newcommand{\killablemutants}{61}
\newcommand{\pointspermutant}{1.6}
\newcommand{\modeldetected}{61}
\newcommand{\modelrate}{100}
\newcommand{\randomrate}{56}
\newcommand{\randomtenxrate}{67}
\newcommand{\handrate}{69}
\newcommand{\equivmutants}{11}
\newcommand{\numpairs}{15}

\newcommand{\maxdivergences}{5}
\newcommand{\equivalenceprofiles}{6}
\newcommand{\httpdeployments}{4}
\newcommand{\httpconforming}{4}
\newcommand{\httpinjected}{12}
\newcommand{\httpinjecteddetected}{12}
\newcommand{\corpusrows}{891}
\newcommand{\zenodocode}{10.5281/zenodo.22244465}

%% file: tab_profiles.tex
\begin{tabular}{lccccccl}
\toprule
Profile & Validity & Captures & Over- & Final- & Void after & Remainder released by \\
        & (days)   & allowed  & capture & capture flag & partial capture & \\
\midrule
Stripe, card default & 7 & 1 & 0\% & yes & no & automatic on partial capture \\
Stripe, multicapture & 7 & 51 & 0\% & yes & no & zero-amount capture with final\_capture=true \\
Stripe, overcapture & 7 & 1 & 15\% & yes & no & automatic on partial capture \\
Adyen, card default & 10 & 1 & 0\% & yes & no & automatic on partial capture \\
Adyen, multiple partial captures & 10 & not fixed & 0\% & no & yes & explicit cancel of the remaining amount \\
Incremental auth (synthetic) & 7 & 2 & 0\% & yes & yes & explicit cancel of the remaining amount \\
\bottomrule
\end{tabular}

%% file: tab_model_check.tex
\begin{tabular}{lrrrrl}
\toprule
Profile & States & Distinct & Depth & Time (s) & Verdict \\
\midrule
Stripe, card default & 182 & 131 & 9 & 0.6 & pass \\
Stripe, multicapture & 1,032 & 642 & 11 & 0.8 & pass \\
Stripe, overcapture & 215 & 155 & 9 & 0.7 & pass \\
Adyen, card default & 249 & 182 & 10 & 0.6 & pass \\
Adyen, multiple partial captures & 1,067 & 738 & 12 & 0.7 & pass \\
Incremental auth (synthetic) & 3,509 & 2,248 & 11 & 0.8 & pass \\
\bottomrule
\end{tabular}

%% file: tab_spec_mutation.tex
\begin{tabular}{llllr}
\toprule
& Change made to the specification & Property targeted & Caught & Trace \\
\midrule
S01 & a final capture no longer checks the authorization deadline & \texttt{NoCaptureAfterExpiry} & yes & 6 \\
S02 & a final capture no longer checks the profile's capture ceiling & \texttt{CaptureWithinLimit} & yes & 5 \\
S03 & the hold may be released again after it has already been released & \texttt{ReleaseAtMostOnce} & yes & 7 \\
S04 & held funds may be released before the hold closes & \texttt{NoReleaseBeforeClose} & yes & 4 \\
S05 & time may pass while a release is overdue & \texttt{BoundedRelease} & yes & 6 \\
S06 & the non-final capture budget is not enforced & \texttt{CaptureCountWithinProfile} & yes & 6 \\
S07 & closing a hold does not record what had been captured & \texttt{NoCaptureAfterClose} & yes & 5 \\
S08 & nothing obliges the provider to ever release the funds & \texttt{EventualRelease} & yes & 7 \\
S09 & nothing obliges the provider to ever expire a stale authorization & \texttt{EventualClose} & yes & 7 \\
S10 & the hold need never be created & \texttt{Termination} & yes & 2 \\
\bottomrule
\end{tabular}

%% file: tab_conformance.tex
\begin{tabular}{lrrrrrrrr}
\toprule
& \multicolumn{2}{c}{Mutants} & \multicolumn{2}{c}{Model suite} & \multicolumn{4}{c}{Fraction detected} \\
\cmidrule(lr){2-3}\cmidrule(lr){4-5}\cmidrule(lr){6-9}
Profile & killable & equiv. & tests & calls & model & random & random$\times$10 & hand \\
\midrule
Stripe, card default & 10 & 2 & 16 & 2,089 & 1.00 & 0.50 & 0.60 & 0.70 \\
Stripe, multicapture & 11 & 1 & 160 & 16,535 & 1.00 & 0.64 & 0.73 & 0.64 \\
Stripe, overcapture & 10 & 2 & 19 & 2,748 & 1.00 & 0.60 & 0.60 & 0.70 \\
Adyen, card default & 10 & 2 & 21 & 3,072 & 1.00 & 0.50 & 0.60 & 0.70 \\
Adyen, multiple partial captures & 10 & 2 & 151 & 20,094 & 1.00 & 0.50 & 0.70 & 0.60 \\
Incremental auth (synthetic) & 10 & 2 & 553 & 82,506 & 1.00 & 0.60 & 0.80 & 0.80 \\
\midrule
\textbf{All profiles} & \textbf{61} &  &  &  & \textbf{1.00} & 0.56 & 0.67 & 0.69 \\
\bottomrule
\end{tabular}

%% file: tab_http.tex
\begin{tabular}{llrrcc}
\toprule
Profile & Shape & Tests & Calls & Clean & Found \\
\midrule
Stripe, card default & stripe & 16 & 2,089 & conforms & 3/3 \\
Stripe, multicapture & stripe & 160 & 16,535 & conforms & 3/3 \\
Adyen, card default & adyen & 21 & 3,072 & conforms & 3/3 \\
Adyen, multiple partial captures & adyen & 151 & 20,094 & conforms & 3/3 \\
\bottomrule
\end{tabular}

%% file: tab_divergences.tex
\begin{tabular}{p{3.0cm}p{2.5cm}p{2.5cm}p{7.4cm}}
\toprule
Call & Left provider & Right provider & Shortest witness \\
\midrule
\multicolumn{4}{l}{\textit{Stripe, card default} (left) vs.\ \textit{Adyen, card default} (right)} \\
advance\_time & status CLOSED & status HELD & \texttt{\scriptsize auth(4) ; wait$\times$3} \\
\addlinespace
\multicolumn{4}{l}{\textit{Stripe, multicapture} (left) vs.\ \textit{Adyen, multiple partial captures} (right)} \\
capture (final), amount 0, 1, 2, 3 & accepted & rejected & \texttt{\scriptsize auth(4) ; cap(1,fin)} \\
capture (non-final), amount 1, 2, 3, 4 & accepted & rejected & \texttt{\scriptsize auth(4) ; cap(4,part)} \\
void & rejected & accepted & \texttt{\scriptsize auth(4) ; cap(1,part) ; void} \\
advance\_time & status CLOSED & status HELD & \texttt{\scriptsize auth(4) ; wait$\times$3} \\
\addlinespace
\multicolumn{4}{l}{\textit{Stripe, overcapture} (left) vs.\ \textit{Adyen, card default} (right)} \\
capture (final), amount 5 & accepted & rejected & \texttt{\scriptsize auth(4) ; cap(5,fin)} \\
advance\_time & status CLOSED & status HELD & \texttt{\scriptsize auth(4) ; wait$\times$3} \\
\bottomrule
\end{tabular}

%% file: holdspec.bbl
\begin{thebibliography}{10}
\providecommand{\url}[1]{#1}
\csname url@samestyle\endcsname
\providecommand{\newblock}{\relax}
\providecommand{\bibinfo}[2]{#2}
\providecommand{\BIBentrySTDinterwordspacing}{\spaceskip=0pt\relax}
\providecommand{\BIBentryALTinterwordstretchfactor}{4}
\providecommand{\BIBentryALTinterwordspacing}{\spaceskip=\fontdimen2\font plus
\BIBentryALTinterwordstretchfactor\fontdimen3\font minus
  \fontdimen4\font\relax}
\providecommand{\BIBforeignlanguage}[2]{{%
\expandafter\ifx\csname l@#1\endcsname\relax
\typeout{** WARNING: IEEEtran.bst: No hyphenation pattern has been}%
\typeout{** loaded for the language `#1'. Using the pattern for}%
\typeout{** the default language instead.}%
\else
\language=\csname l@#1\endcsname
\fi
#2}}
\providecommand{\BIBdecl}{\relax}
\BIBdecl

\bibitem{sharma2026idempotency}
A.~Sharma, ``A provider-aware formal contract and conformance suite for payment
  idempotency,'' 2026, manuscript under review.

\bibitem{stripe2026hold}
{Stripe}, ``Place a hold on a payment method,''
  \url{https://docs.stripe.com/payments/place-a-hold-on-a-payment-method},
  2026, accessed 1 September 2026.

\bibitem{stripe2026multicapture}
------, ``Capture a payment multiple times,''
  \url{https://docs.stripe.com/payments/multicapture}, 2026, accessed 1
  September 2026.

\bibitem{stripe2026overcapture}
------, ``Capture more than the authorized amount on a payment,''
  \url{https://docs.stripe.com/payments/overcapture}, 2026, accessed 1
  September 2026.

\bibitem{adyen2026capture}
{Adyen}, ``Capture,'' \url{https://docs.adyen.com/online-payments/capture/},
  2026, accessed 1 September 2026.

\bibitem{adyen2026adjust}
------, ``Adjust authorisation,''
  \url{https://docs.adyen.com/online-payments/adjust-authorisation/}, 2026,
  accessed 1 September 2026.

\bibitem{abadi1994oldfashioned}
M.~Abadi and L.~Lamport, ``An old-fashioned recipe for real time,'' \emph{ACM
  Transactions on Programming Languages and Systems}, vol.~16, no.~5, pp.
  1543--1571, 1994.

\bibitem{howard2025ccf}
H.~Howard, M.~A. Kuppe, E.~Ashton, A.~Chamayou, and N.~Crooks, ``Smart casual
  verification of the confidential consortium framework,'' in \emph{22nd USENIX
  Symposium on Networked Systems Design and Implementation (NSDI 25)}.\hskip
  1em plus 0.5em minus 0.4em\relax USENIX Association, 2025, arXiv:2406.17455.

\bibitem{chow1978testing}
T.~S. Chow, ``Testing software design modeled by finite-state machines,''
  \emph{IEEE Transactions on Software Engineering}, vol. SE-4, no.~3, pp.
  178--187, 1978.

\bibitem{lee1996principles}
D.~Lee and M.~Yannakakis, ``Principles and methods of testing finite state
  machines --- a survey,'' \emph{Proceedings of the IEEE}, vol.~84, no.~8, pp.
  1090--1123, 1996.

\bibitem{demillo1978hints}
R.~A. DeMillo, R.~J. Lipton, and F.~G. Sayward, ``Hints on test data selection:
  Help for the practicing programmer,'' \emph{Computer}, vol.~11, no.~4, pp.
  34--41, 1978.

\bibitem{jia2011analysis}
Y.~Jia and M.~Harman, ``An analysis and survey of the development of mutation
  testing,'' \emph{IEEE Transactions on Software Engineering}, vol.~37, no.~5,
  pp. 649--678, 2011.

\bibitem{basin2021emv}
D.~Basin, R.~Sasse, and J.~Toro-Pozo, ``The {EMV} standard: Break, fix,
  verify,'' in \emph{2021 IEEE Symposium on Security and Privacy (SP)}.\hskip
  1em plus 0.5em minus 0.4em\relax IEEE, 2021, pp. 1766--1781,
  arXiv:2006.08249.

\bibitem{grundmann2023lightning}
M.~Grundmann and H.~Hartenstein, ``Towards a formal verification of the
  lightning network with {TLA+},'' \emph{arXiv preprint arXiv:2307.02342},
  2023.

\bibitem{newcombe2014amazon}
C.~Newcombe, T.~Rath, F.~Zhang, B.~Munteanu, M.~Brooker, and M.~Deardeuff,
  ``Use of formal methods at {Amazon} web services,'' in \emph{Communications
  of the ACM}, vol.~58, no.~4, 2015, pp. 66--73.

\bibitem{godefroid2020differential}
P.~Godefroid, D.~Lehmann, and M.~Polishchuk, ``Differential regression testing
  for {REST} {APIs},'' in \emph{Proceedings of the 29th ACM SIGSOFT
  International Symposium on Software Testing and Analysis (ISSTA)}.\hskip 1em
  plus 0.5em minus 0.4em\relax ACM, 2020, pp. 312--323.

\end{thebibliography}
